\documentclass[12pt]{iopart}

\usepackage{epsfig,graphics}
\begin{document}
\title[Detailed electron and positron capture rates on $^{24}$Mg] {Detailed
microscopic calculation of stellar electron and positron capture
rates on $^{24}$Mg for O+Ne+Mg core simulations}
\author{Jameel-Un Nabi}
\address{Faculty of Engineering Sciences, GIK Institute of Engineering Sciences and
Technology, Topi 23640, NWFP, Pakistan \\ Current Address: ICTP,
Strada Costiera 11, 34014, Trieste, Italy} \ead{jnabi00@gmail.com}
\begin{abstract}
Few white dwarfs, located in binary systems, may acquire
sufficiently high mass accretion rates resulting in the burning of
carbon and oxygen under nondegenerate conditions forming a O+Ne+Mg
core. These O+Ne+Mg cores are gravitationally less bound than more
massive progenitor stars and can release more energy due to the
nuclear burning. They are also amongst the probable candidates for
low entropy r-process sites. Recent observations of subluminous Type
II-P supernovae (e.g., 2005cs, 2003gd, 1999br, 1997D) were able to
rekindle the interest in 8 -- 10 M$_{\odot}$ which develop O+Ne+Mg
cores. Microscopic calculations of capture rates on $^{24}$Mg, which
may contribute significantly to the collapse of O+Ne+Mg cores, using
shell model and proton-neutron quasiparticle random phase
approximation (pn-QRPA) theory, were performed earlier and
comparisons made. Simulators, however, may require these capture
rates on a fine scale. For the first time a detailed microscopic
calculation of the electron and positron capture rates on $^{24}$Mg
on an extensive temperature-density scale is presented here. This
type of scale is more appropriate for interpolation purposes and of
greater utility for simulation codes. The calculations are done
using the pn-QRPA theory using a separable interaction. The
deformation parameter, believed to be a key parameter in QRPA
calculations, is adopted from experimental data to further increase
the reliability of the QRPA results. The resulting calculated rates
are up to a factor of 14 or more enhanced as compared to shell model
rates and may lead to some interesting scenario for core collapse
simulators.

\end{abstract}
 \maketitle

\section{Introduction}
White dwarfs located in a binary system may end their lives in two
possible ways. They may accrete from a companion and achieve the
Chandrasekhar mass thereby triggering a thermonuclear runaway of the
object and ultimately exploding as a Type Ia supernova. In this case
no remnant is left behind. Alternatively these massive white dwarfs,
for sufficiently high mass accretion rates, may allow the formation
of O+Ne+Mg cores and due to the high prevailing central density
(beyond $10^{10} g cm^{-3}$) experience rapid electron capture that
lead to the collapse of the core [1]. This is termed as
accretion-induced collapse (AIC). The end product is a neutron star
(in some double-degenerate scenario, two white dwarfs in a
short-period binary system may eventually coalesce to form a massive
white dwarf that exceeds the Chandrasekhar mass limit and in this
case transition to a black hole is possible if the total
proto-neutron star mass exceeds the general-relativistic limit for
gravitational stability [2]). The ultimate fate of these white
dwarfs are dependent on many factors, e.g. temperature of the
environment, the mass accretion rate on the newly formed white
dwarf, and the mass of each partner white dwarf [2]. The occurrence
rate of the AIC of white dwarfs is not determined reliably, and are
not expected to occur more than once per 20 -- 50 standard Typa Ia
events (see for example [2]).

Supernova explosions from the collapse of O+Ne+Mg cores remain a
potential research interest for astrophysical reasons. Its smaller
core and smaller gravitational potential may allow the star to
explode hydrodynamically. They are also proposed as a probable
candidate for the site of r-process as compared to the "neutrino
wind" mechanism based on the condition that they may explode by the
prompt bounce-shock mechanism [2-5]. A detailed analysis of the core
evolution of massive stars may be found in [1].

The evolution of the stars in the mass range 8 -- 10 M$_{\odot}$
develops central cores which are composed of $^{16}$O, $^{20}$Ne and
$^{24}$Mg. Oda et al. [6] employed shell model wave functions of the
$sd$-shell nuclei developed by Wildenthal [7] and calculated the
capture rates which contribute to the collapse of the O+Ne+Mg core.
Oda et al. [6] pointed out three different series of electron
capture in the O+Ne+Mg core of the 8 -- 10 M$_{\odot}$ stars and
placed them in the order of low threshold energy as $^{24}$Mg
$\rightarrow$ $^{24}$Na $\rightarrow$ $^{24}$Ne, $^{20}$Ne
$\rightarrow$ $^{20}$F $\rightarrow$ $^{20}$O, and $^{16}$O
$\rightarrow$ $^{16}$N $\rightarrow$ $^{16}$C. They, however,
renounced the last series in their calculations because of its high
threshold energy which did not contribute significantly to the
initiation of the collapse of the O+Ne+Mg core of the 8 -- 10
M$_{\odot}$ stars.

A reasonably large number of recent observations and accumulation of
data regarding Type II-P supernovae lead to a rejuvenation of debate
on the fate of O+Ne+Mg cores. Earlier Guti\'{e}rrez et al. [8]
argued that the abundance of $^{24}$Mg was considerably reduced in
updated evolutionary calculations. However the procedure adopted by
them was not fully consistent as they kept the ratio of oxygen to
neon constant while parameterizing the abundance of $^{24}$Mg. Later
Kitaura et al. [9] presented simulation result of these cores
(keeping capture rates on $^{24}$Mg as a key ingredient) using an
improved neutrino transport treatment. Their outcome was not a
prompt but a delayed explosion. Kitaura and collaborators [9] used
the shell model capture rates of Takahara et al. [10] as a key
nuclear input parameter in their simulation codes in the non-nuclear
statistical equilibrium regime. One of the reasons for this
spherical core collapse simulations not to explode was the reduced
electron capture rates employed in their simulations. This reduced
capture rate slowed the collapse process and resulted in a large
shock radius.

 Electron captures on proton and positron captures on neutron play a
very crucial role in the supernovae dynamics. During the collapse
and accretion phases, these processes exhaust electrons, thus
decreasing the degenerate pressure of electrons in the stellar core.
Meanwhile, they produce neutrinos which carry the binding energy
away. Therefore, electron and positron captures play key role in the
dynamics of the formation of bounce shock of supernova. The Type II
supernovae take place in heavy stars. The positron captures are of
great importance in high temperature and low density locations. In
such conditions, a rather high concentration of positron can be
reached from $e^{-} +e^{+} \leftrightarrow \gamma +\gamma $
equilibrium favoring the $e^{-} e^{+} $ pairs. The electron captures
on proton and positron captures on neutron are considered important
ingredients in the modeling of Type II supernovae [11].

The key purpose of reporting this work is the presentation of the
newly calculated microscopic calculation of the electron and
positron capture rates using the pn-QRPA theory with separable
interaction. Due to the extreme conditions prevailing in the cores
of O+Ne+Mg stars, interpolation of calculated rates within large
intervals of temperature-density points posed some uncertainty in
the values of capture rates for collapse simulators. As such the
calculation is done on a detailed temperature and density grid
pertinent to presupernova and supernova environment and should
prove more suitable for running on simulation codes.

The pn-QRPA theory is proven to be quite successful in the
calculation of stellar weak rates. Earlier Nabi and Klapdor [12]
used the pn-QRPA theory to calculate the weak interaction rates for
178 $sd$-shell nuclei [12] and 650 fp/fpg-shell [13] nuclide in
stellar matter. The reliability of the pn-QRPA calculations was
discussed in detail by Nabi and Klapdor [13]. There the authors
compared the measured data (half lives and Gamow-Teller (GT)
strength) of thousands of nuclide with the pn-QRPA calculations and
got fairly good comparison. Later the decay and capture rates of
nuclei of astrophysical importance were studied separately in detail
and were compared with earlier calculations wherever possible both
in sd-shell [14] and fp-shell (e.g. [15-17]) regions. Compared to
shell model calculations, the pn-QRPA gives similar accuracy in
reproducing beta-decay and capture rates in $sd$-shell nuclei [14].
The deformation parameter is as an important parameter for QRPA
calculations as pairing [18]. For the case of even-even nuclei,
experimental deformations are available [19] and were employed in
this work.

In this paper a detailed calculation of electron and positron
capture rates on $^{24}$Mg is being presented for the first time
at temperature-density intervals suitable for simulation purposes.
The presence of $^{24}$Mg in the core is a result of the previous
phase of carbon burning and its relevance is due to its lower
electron capture threshold. Section 2 briefly discusses the
formalism of the pn-QRPA calculations. Section 3 presents some of
the calculated results. Comparison with earlier calculations is
also included in this section. The conclusions are given in
Section 4 and at the end Table 1 presents the detailed calculation
of electron and positron capture rates on $^{24}$Mg.

\section{Formalism}
The Hamiltonian for the calculations was chosen to be of the form
\begin{equation}
H^{QRPA} =H^{sp} +V^{pair} +V_{GT}^{ph} +V_{GT}^{pp},
\end{equation}
where $H^{sp}$ is the single-particle Hamiltonian, $V^{pair}$  is
the pairing force, $V_{GT}^{ph}$ is the particle-hole (ph)
Gamow-Teller force, and $V_{GT}^{pp}$  is the particle-particle (pp)
Gamow-Teller force. Single particle energies and wave functions were
calculated in the Nilsson model, which takes into account nuclear
deformations. Pairing was treated in the BCS approximation. The
proton-neutron residual interactions occurred as particle-hole and
particle-particle interaction. The interactions were given separable
form and were characterized by two interaction constants $\chi$  and
$\kappa$, respectively. In this work, the values of $\chi$ and
$\kappa$ was taken as 0.001 MeV and 0.05 MeV, respectively. Other
parameters required for the calculation of weak rates are the
Nilsson potential parameters, the deformation, the pairing gaps, and
the Q-value of the reaction. Nilsson-potential parameters were taken
from Ref. [20] and the Nilsson oscillator constant was chosen as
$\hbar \omega=41A^{-1/3}(MeV)$ (the same for protons and neutrons).
The calculated half-lives depend only weakly on the values of the
pairing gaps [21]. Thus, the traditional choice of $\Delta _{p}
=\Delta _{n} =12/\sqrt{A} (MeV)$ was applied in the present work.
The deformation parameter is recently argued to be one of the most
important parameters in QRPA calculations [18] and as such rather
than using deformations from some theoretical mass models (as used
in earlier calculations of pn-QRPA capture rates) the experimentally
adopted value of the deformation parameters for $^{24}$Mg, extracted
by relating the measured energy of the first $2^{+}$ excited state
with the quadrupole deformation, was taken from Raman et al. [19].
Q-values were taken from the recent mass compilation of Audi et al.
[22].

The electron capture (ec) and positron capture (pc) rates of a
transition from the $ith$ state of the parent to the $jth$ state
of the daughter nucleus is given by
\begin{equation}
\lambda ^{^{ec(pc)} } _{ij} =\left[\frac{\ln 2}{D}
\right]\left[f_{ij} (T,\rho ,E_{f} )\right]\left[B(F)_{ij}
+\left({\raise0.7ex\hbox{$ g_{A}  $}\!\mathord{\left/ {\vphantom
{g_{A}  g_{V} }} \right.
\kern-\nulldelimiterspace}\!\lower0.7ex\hbox{$ g_{V}  $}}
\right)^{2} B(GT)_{ij} \right].
\end{equation}
The value of D was taken to be 6295s [23] and the ratio of the axial
vector to the vector coupling constant as -1.254 [24]. $B_{ij}'s$
are the sum of reduced transition probabilities of the Fermi B(F)
and GT transitions B(GT). The $f_{ij}'s$ are the phase space
integrals. Details of the calculations of phase space integrals,
reduced transition probabilities and choice of Gamow-Teller strength
parameters can be found in [12,13,25].

The total electron (positron) capture rate per unit time per
nucleus was then calculated using
\begin{equation}
\lambda^{ec(pc)} =\sum _{ij}P_{i} \lambda _{ij}^{ec(pc)}.
\end{equation}
The summation over all initial and final states was carried out
until satisfactory convergence in the rate calculations was
achieved. Here $P_{i}$ is the probability of occupation of parent
excited states and follows the normal Boltzmann distribution.

It is pertinent to mention that experimental data was incorporated
in the calculations wherever possible to further increase the
reliability of the calculated capture rates. In addition to the
incorporation of the measured value of the deformation parameter,
the calculated excitation energies (along with their log $ft$
values) were replaced with an experimental one when they were within
0.5 MeV of each other. Missing measured states were inserted and
inverse and mirror transitions were also taken into account. No
theoretical levels were replaced with the experimental ones beyond
the excitation energy for which experimental compilations have no
definite spin and/or parity. This means that no theoretical levels
were replaced in the parent $^{24}$Mg beyond 9.3 MeV. For the
daughter nuclei no replacement was done in $^{24}$Na beyond 1.3 MeV
(for the electron capture direction) and in $^{24}$Al beyond 1.1 MeV
(for the positron capture direction).

\section{Results and Discussions}
The Gamow-Teller strength distributions, B(GT$\pm$), and the
associated electron and positron capture rates on $^{24}$Mg were
calculated using the pn-QRPA theory. Quenching of the GT strength
was taken into account and a standard quenching factor of 0.77 was
used [26]. The calculation was performed for a total of 132 excited
states of $^{24}$Mg up to excitation energies in the vicinity of 40
MeV (corresponding to first 100 excited states) in the daughters
$^{24}$Na and $^{24}$Al.

The summed B(GT+) strength is shown in Fig.1. It is to be noted that
the experimentally extracted total B(GT+) strength for excitation
energy region up to 7 MeV (where the density of states is still low
enough to analyze single peaks) is 1.36 [27]. The corresponding
calculated value reported in this work comes to be 2.53. The earlier
reported pn-QRPA value was 2.65 [14]. The differences in the BGT(+)
value and in the calculated capture rates compared to those of Ref.
[14] are attributed primarily to the choice of deformation parameter
and incorporation of all experimental levels (along with their $log
ft$ values, if available) as stated in previous section. Whereas the
authors in [14] incorporated a deformation parameter of 0.435
according to the mass formula of M\"{o}ller and Nix [28], in this
work the measured value of 0.606 [19] is taken to be the deformation
parameter for $^{24}$Mg. Details on the sensitivity of B(GT+)
strength as a function of deformation parameter can be found in
[29]. Regarding the measured value of the GT strength, it is to be
noted that Rakers and collaborators [27] employed a $sd$-model space
and the universal $sd$ residual interaction [30,31] to calculate the
wave functions. The authors in [27] argued that the full model space
calculation was not possible and they had to truncate the model
space which subsequently lead to uncertainties and the authors had
some reservations in the interpretation of their data. The pn-QRPA
value reported in this work is close to the reported shell model
value of 2.1 of Brown and Wildenthal [30] and Wildenthal [31].
Takahara et al. [10] reported a value of 1.30. The total calculated
GT strength in this work is 3.34 (quenched).

Fig.2 depicts the summed B(GT-) strength as a function of
excitation energies in the daughter nucleus $^{24}$Al. The total
strength comes to 3.34 (quenched) and the Ikeda sum rule is
fulfilled in the calculations.

Moving on to the calculation of capture rates, Fig.3 shows four
panels depicting the calculated electron capture rates at selected
temperature and density domain. The upper left panel shows the
electron capture rates in low-density region ($\rho [gcm^{-3}]
=10^{0.5}, 10^{1.5}$ and $10^{2.5}$), the upper right in
medium-low density region ($\rho [gcm^{-3}] =10^{3.5}, 10^{4.5}$
and $10^{5.5}$), the lower left in medium-high density region
($\rho [gcm^{-3}] =10^{6.5}, 10^{7.5}$ and $10^{8.5}$) and finally
the lower right panel depicts the calculated electron capture
rates in high density region ($\rho [gcm^{-3}] =10^{9.5},
10^{10.5}$ and $10^{11}$). The capture rates are given in
logarithmic scale. It can be seen from this figure that in the low
density region the capture rates, as a function of temperature,
are more or less superimposed on one another. This means that
there is no appreciable change in the rates when increasing the
density by an order of magnitude. However as one moves from the
medium low density region to high density region these rates start
to 'peel off' from one another. Orders of magnitude difference in
rates are observed (as a function of density) in high density
regions. For a given density the rates increase monotonically with
increasing temperatures. One also notices that the electron
capture rates coincide at T$_{9}$ = 30K except for the high
density region (where T$_{9}$ gives the stellar temperature in
units of $10^{9}$K).

One of the channels for the energy release from the star is the
neutrino emission which is mainly from the $e/e^{+}$ capture on
nucleons and $e^{\pm}$ annihilation. Positron capture plays a
crucial role in the dynamics of stellar core. They play an
indirect role in the reduction of degeneracy pressure of the
electrons in the core. Fig.4 shows four panels depicting the
calculated positron capture rates at selected temperature and
density domain. The upper left panel shows the positron capture
rates in low-density region ($\rho [gcm^{-3}] =10^{0.5}, 10^{1.5}$
and $10^{2.5}$), the upper right in medium-low density region
($\rho [gcm^{-3}] =10^{3.5}, 10^{4.5}$ and $10^{5.5}$), the lower
left in medium-high density region ($\rho [gcm^{-3}] =10^{6.5},
10^{7.5}$ and $10^{8.5}$) and finally the lower right panel
depicts the calculated positron capture rates in high density
region ($\rho [gcm^{-3}] =10^{9.5}, 10^{10.5}$ and $10^{11}$). The
positron capture rates are given in logarithmic scales. It is to
be noted that the positron capture rates are very slow as compared
to electron capture on $^{24}$Mg. The positron capture rates
enhance as temperature of the stellar core increases and decrease
with increasing stellar density. One should note the order of
magnitude differences in positron capture rates as the stellar
temperature increases. It can be seen from this figure that in the
low density region the positron capture rates, as a function of
temperature, are more or less superimposed on one another (similar
to the case of electron capture rates). One also observes that the
positron capture rates are almost the same for the densities in
the range $(10-10^{6})g/cm^{3}$. However as one moves from the
medium high density region to high density region these rates also
start to 'peel off' from one another. When the densities increase
beyond the above stated range a decline in the positron capture
rate starts. For a given density the rates increase monotonically
with increasing temperatures.

The reported capture rates on $^{24}$Mg were also compared against
two earlier calculations. Fuller, Fowler and Newman [32] (hereafter
FFN) compiled the experimental data and calculated electron and
positron capture rates (besides other weak-interaction mediated
rates) for the nuclei in the mass range A = 21-60 for an extended
grid of temperature and density. The GT strength and excitation
energies were calculated using a zero-order shell model. For the
discrete transitions for which the $ft$ values were not available
FFN took log $ft$ = 5.0. Later Oda et al. [6] performed an extensive
calculation of stellar weak interaction rates of $sd$-shell nuclei
in full (sd)$^{n}$-shell model space. They used the effective
interaction of Wildenthal [7,33,34] and the available experimental
compilations for their calculations.

Fig.5 compares the reported electron capture rates against those of
Oda et al. [6] and FFN [32]. It is to be noted that the first two
factors in the calculation of capture rates in Eqtn. (2) are similar
in all three calculations. Whereas the first factor is a constant,
the second factor constitutes the calculation of phase space
integrals. The formalism for the calculation of these integrals is
similar in all calculations and are dependent only on corresponding
stellar temperatures, densities and Fermi energies. The main
difference in the three calculations arises because of the third
factor (calculation of reduced transition probabilities) which
contains in it the in-built nuclear structure effects. In Fig.5
$R_{ec}(QRPA/OHMTS)$ (in the upper panel) is the ratio of the
reported electron capture rates to those calculated using shell
model by Oda et al [6] whereas $R_{ec}(QRPA/FFN)$ (in the lower
panel) is the ratio of the reported electron capture rates to those
calculated by FFN [32]. The shell model electron capture rates are
usually in good agreement with the FFN rates. The reported pn-QRPA
electron capture rates are enhanced for all values of temperature
and density compared to both shell model and FFN results. This is
one major difference with the earlier reported capture rates on
$^{24}$Mg [14]. For densities around $\rho [gcm^{-3}] =10^{3}$, the
pn-QRPA rates are enhanced by as much as factor of 14 or more as
compared to the corresponding shell model rates. As the density of
the stellar matter increases, the enhancement ratio decreases. The
pn-QRPA electron capture rates are enhanced by as much as factor of
13 around $\rho [gcm^{-3}] =10^{7}$ and roughly by a factor of 4 at
high density around $\rho [gcm^{-3}] =10^{11}$. Core-collapse
simulators should take note of these enhanced reported electron
capture rates. These rates might contribute to some interesting
simulation results. A similar comparison for positron capture (pc)
rates against previous calculations is shown in the two panels of
Fig.6. Here one notes that the reported pn-QRPA positron capture
rates are relatively in better agreement with the FFN rates although
both FFN and shell model rates are enhanced. Due to orders of
magnitude differences the ratios are plotted on a logarithmic scale
in Fig.6. The comparisons are fairly good at high temperatures. It
is worth mentioning that the positron capture rates are very small
numbers and can change by orders of magnitude by a mere change of
0.5 MeV, or less, in parent or daughter excitation energies and are
more reflective of the uncertainties in the calculation of energies.

The calculated electron and positron capture rates on $^{24}$Mg on a
fine scale of stellar temperature-density is given in Table 1. The
calculated rates are tabulated in logarithmic (to base 10) scale. In
the table, -100.000 means that the rate is smaller than 10$^{-100}$.
The first column gives log($\rho Y_{e}$) in units of $g cm^{-3}$,
where $\rho$ is the baryon density and $Y_{e}$ is the ratio of the
electron number to the baryon number. Stellar temperatures ($T_{9}$)
are measured in $10^{9}$ K. Stated also are the values of the Fermi
energy of electrons in units of MeV. $\lambda^{ec}$($\lambda^{pc}$)
are the electron(positron) capture rates in units of  $sec^{-1}$
(Eqtn. 3). The ASCII file of Table 1 is also available and can be
received from the author upon request.

\section{Conclusions}
The pn-QRPA theory with separable interaction was used to calculate
the electron and positron capture rates on $^{24}$Mg on a much
detailed temperature-density grid point suitable for simulation
purposes. Deformation parameter, which is believed to be one of the
important parameters of pn-QRPA calculations, was taken from the
experimental compilation to further increase the reliability of the
calculations. The reported electron capture rates are enhanced, by
as much as a factor of 12, compared to the shell model results of
Oda and collaborators [6].

There exists a wide variety of discrepant results in the supernova
simulations of O+Ne+Mg cores including prompt explosions, delayed
explosions and no explosions. Fryer et al. [35] and Kitaura et al.
[9] pointed that a possible explanation to these varying results
could be explained by the different nuclear equation of states
employed in the simulations and the treatment of capture rates on a
large variety of nuclei both in nuclear statistical equilibrium
(NSE) and in the non-NSE regime. Recent simulations of O+Ne+Mg cores
by Guti\'{e}rrez et al. [8] and Kitaura et al. [9] employ the shell
model capture rates of Oda et al. [6] and Takahara et al. [10],
respectively. The spherical core collapse simulations do not explode
partly because of the reduced electron capture rates in the non-NSE
regime slowing the collapse and resulting in a large shock radius.

It can be expected that a parameter-free multi-dimensional model,
with neutrino transport included consistently throughout the
entire mass, including a complete treatment of multidimensional
convection and burning phases might lead to a better understanding
of the phenomena of supernova explosions. Incorporation of the
reported (enhanced) capture rates might point toward still lower
values of $Y_{e}$ and lower entropy in the stellar core and might
give some interesting results for simulators transforming a
collapse into an explosion.

\ack The author would like to acknowledge the local hospitality
provided by the Abdus Salam ICTP, Trieste, where part of this
project was completed.

\newpage
\begin{figure}[htbp]
\begin{center}
\includegraphics[width=0.8\textwidth]{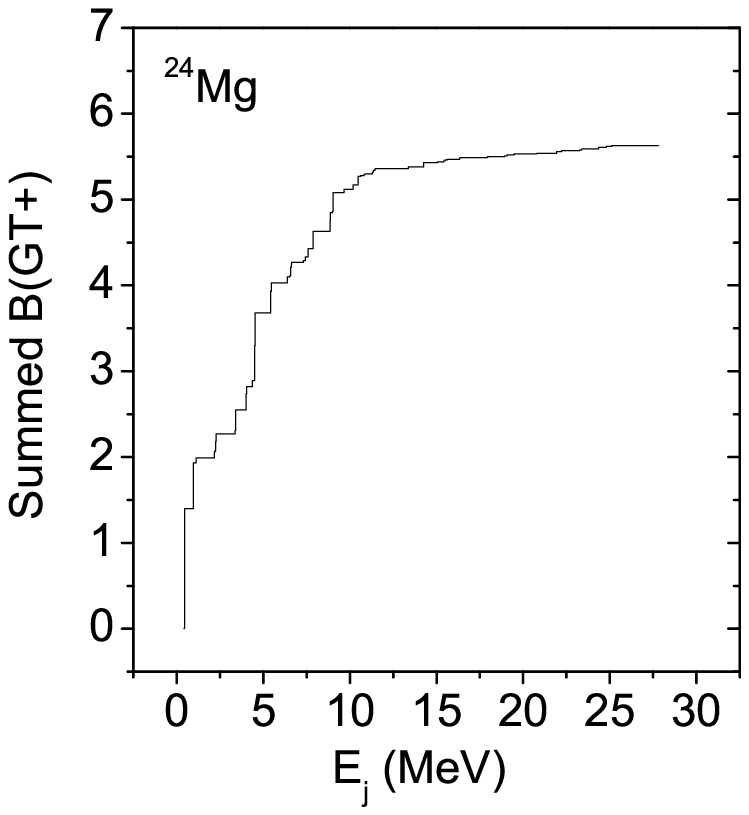}
\caption{Cumulative sum of the unquenched B(GT$^{+}$) values.
$E_{j}$ represents the daughter excitation energies in $^{24}$Na.
}\label{figure1}
\end{center}
\end{figure}
\begin{figure}[htbp]
\begin{center}
\includegraphics[width=0.8\textwidth]{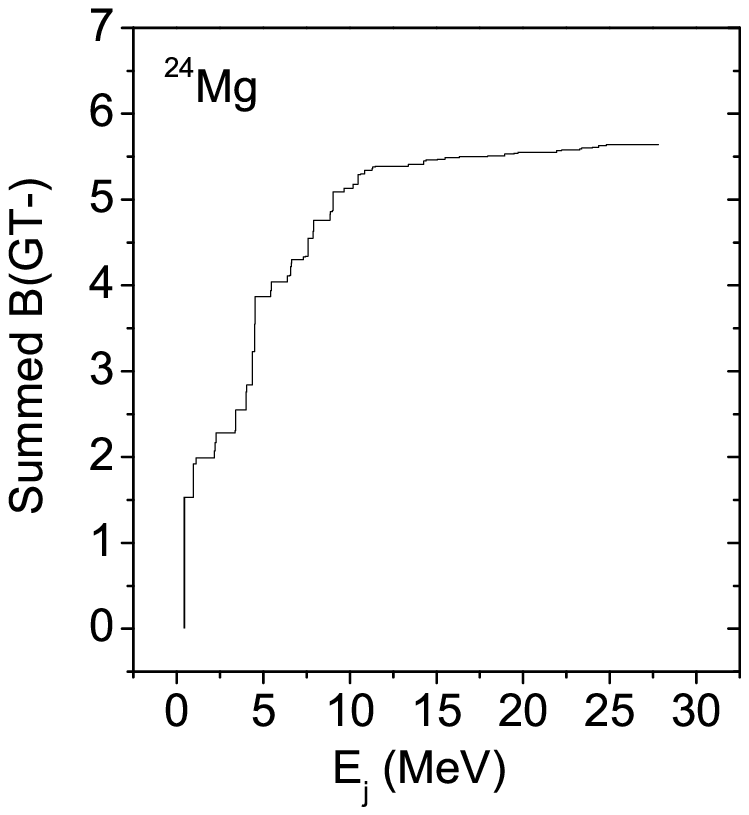}
\caption{Cumulative sum of the unquenched B(GT$^{-}$) values.
$E_{j}$ represents the daughter excitation energies in $^{24}$Al.
}\label{figure2}
\end{center}
\end{figure}
\begin{figure}[htbp]
\begin{center}
\includegraphics[width=0.8\textwidth]{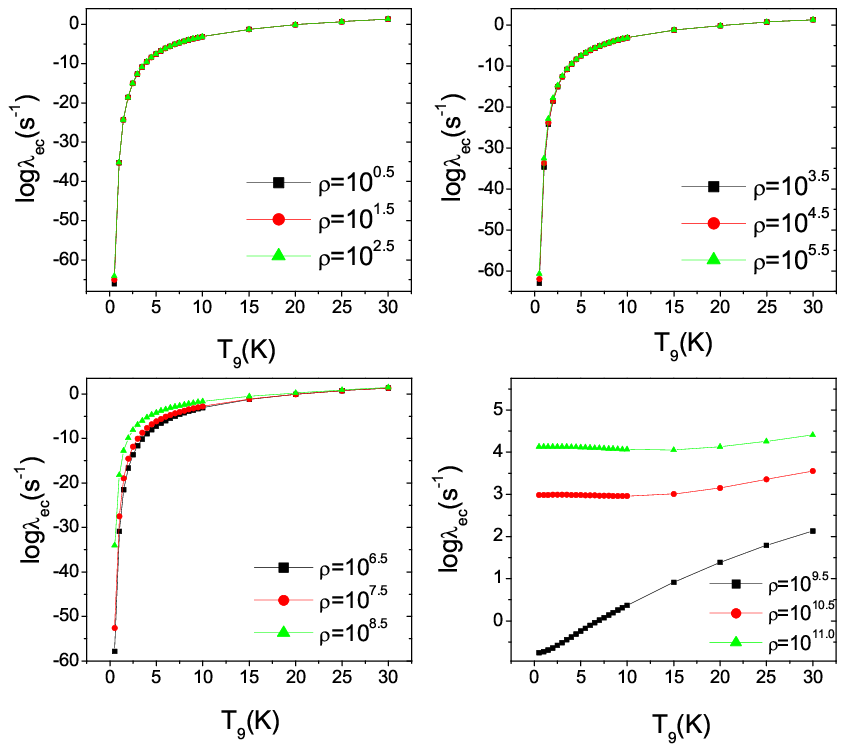}
\caption{(Color online) Electron capture rates on $^{24}$Mg, as
function of temperature, for different selected densities .
Densities are in units of $gcm^{-3}$. Temperatures are measured in
$10^{9}$ K and log$\lambda_{ec}$ represents the log of electron
capture rates in units of $sec^{-1}$.}\label{figure3}
\end{center}
\end{figure}
\begin{figure}[htbp]
\begin{center}
\includegraphics[width=0.8\textwidth]{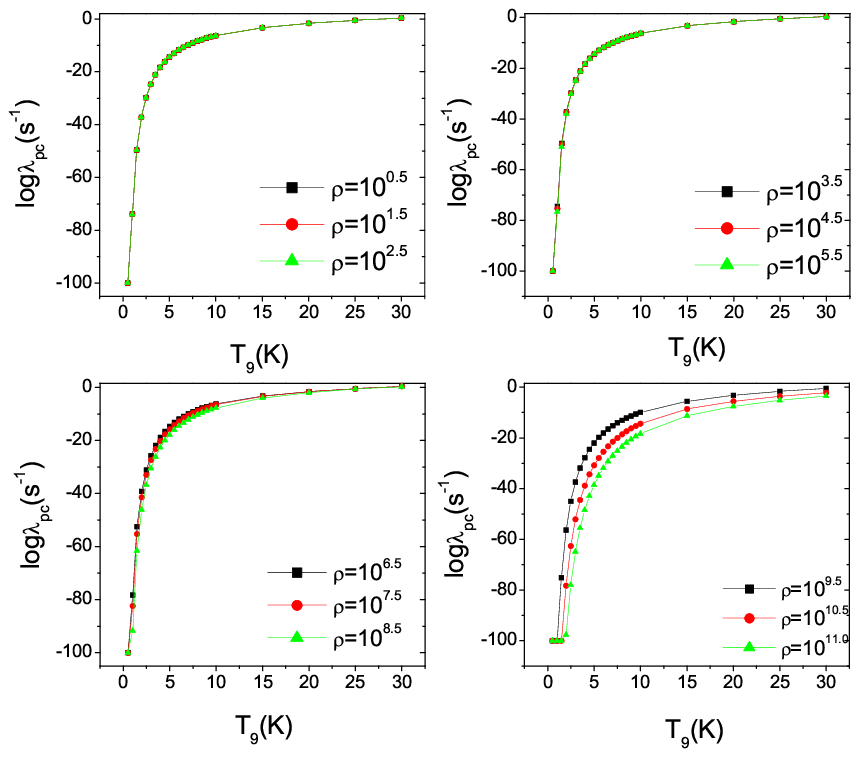}
\caption{(Color online) Positron capture rates on $^{24}$Mg, as
function of temperature, for different selected densities .
Densities are in units of $gcm^{-3}$. Temperatures are measured in
$10^{9}$ K and log$\lambda_{pc}$ represents the log of positron
capture rates in units of $sec^{-1}$.}\label{figure4}
\end{center}
\end{figure}
\begin{figure}[htbp]
\begin{center}
\includegraphics[width=1.2\textwidth]{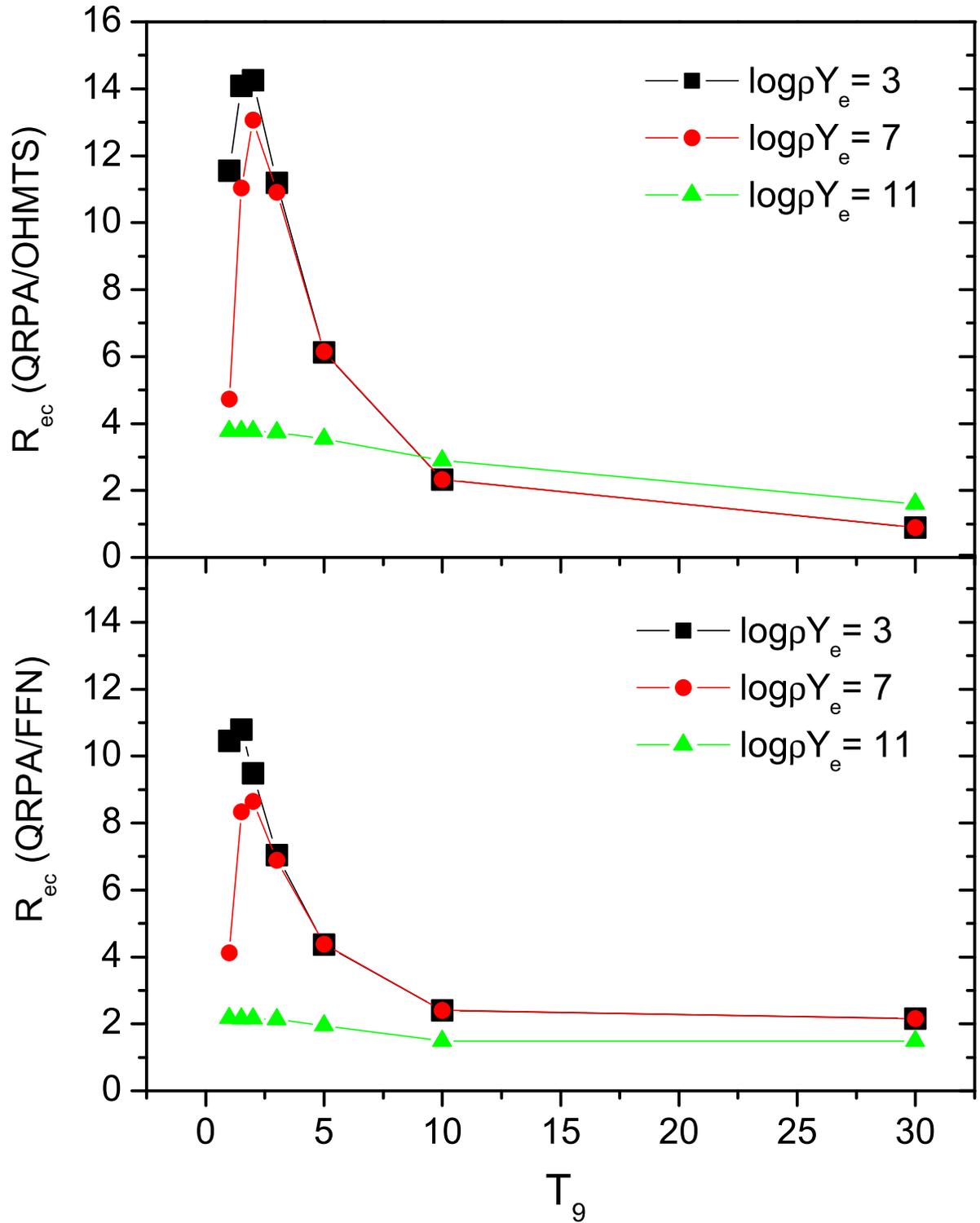}
\caption{(Color online) Ratios of reported electron capture rates to
those calculated using shell model [6] (upper panel) and FFN [32]
(lower panel) on $^{24}$Mg as function of stellar temperatures and
densities. T$_{9}$ measures the stellar temperature in units of
$10^{9}$K.}\label{figure5}
\end{center}
\end{figure}
\begin{figure}[htbp]
\begin{center}
\includegraphics[width=1.2\textwidth]{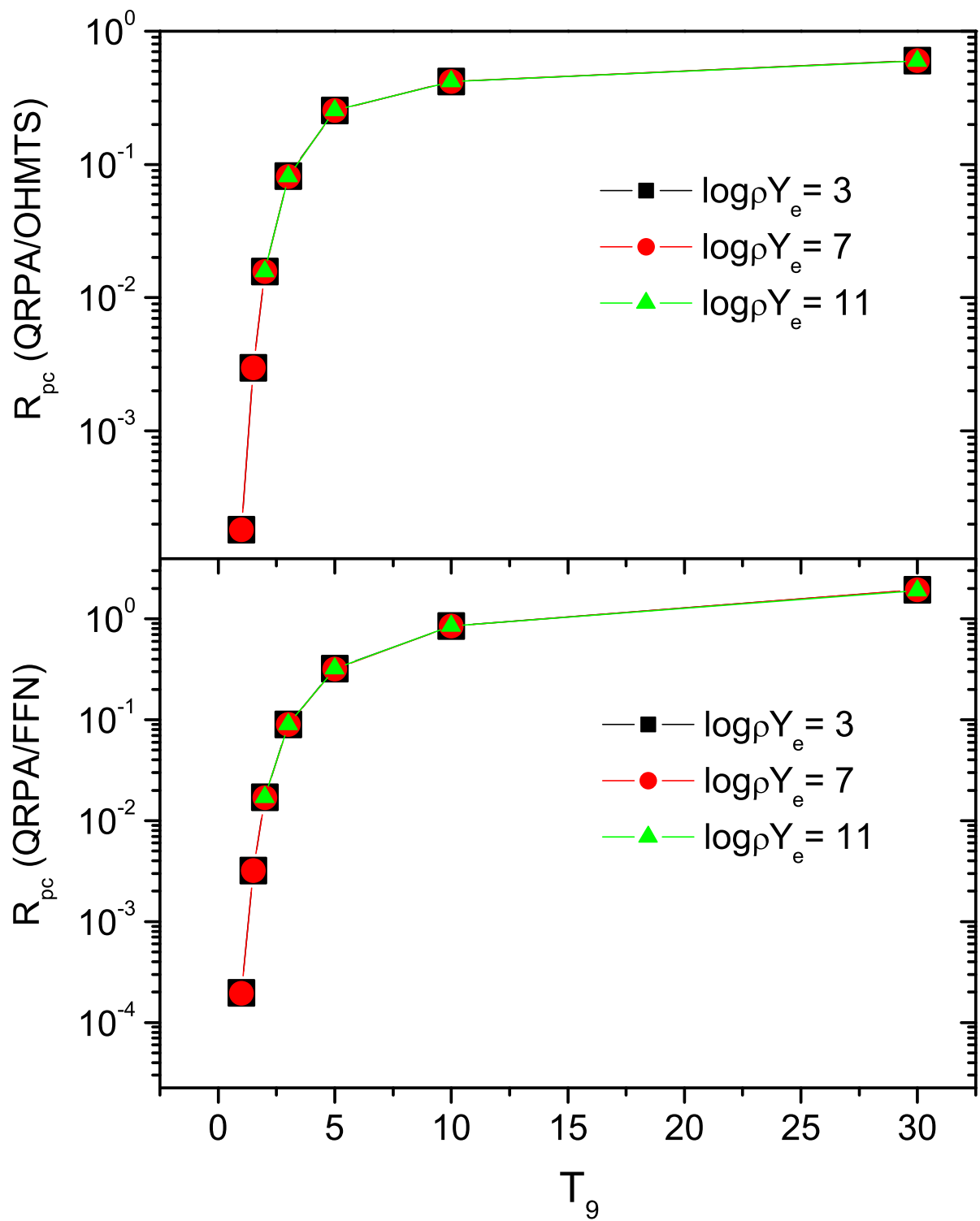}
\caption{(Color online) Same as Fig. 5 but for positron capture (pc)
rates.}\label{figure6}
\end{center}
\end{figure}

\clearpage \textbf{Table 1:} Calculated electron and positron
capture rates on $^{24}$Mg for different selected densities and
temperatures in stellar matter. log($\rho Y_{e}$) has units of $g
cm^{-3}$, where $\rho$ is the baryon density and $Y_{e}$ is the
ratio of the electron number to the baryon number. Temperatures
($T_{9}$) are measured in $10^{9}$ K. $E_{f}$ is the total Fermi
energy of electrons including the rest mass(MeV).
$\lambda^{ec}$($\lambda^{pc}$) are the electron(positron) capture
rates $(sec^{-1})$. The calculated rates are tabulated in
logarithmic (to base 10) scale.
In the table, -100.000 means that the rate is smaller than 10$^{-100}$. \\
\scriptsize{
\begin{center}
\begin{tabular}{ccccc|ccccc|ccccc} \\ \hline
$log\rho Y_{e}$& $T_{9}$ & $E_{f}$ & $\lambda^{ec}$ &
$\lambda^{pc}$ & $log\rho Y_{e}$& $T_{9}$ & $E_{f}$ &
$\lambda^{ec}$ & $\lambda^{pc}$ & $log\rho Y_{e}$& $T_{9}$ &
$E_{f}$ & $\lambda^{ec}$ & $\lambda^{pc}$   \\\hline
0.5& 0.5& 0.065& -66.044& -100&  1&   8.5& 0&   -3.952&  -7.796&  2&   4.5& 0&   -8.4&    -16.178\\
0.5& 1  & 0   &-35.225& -73.878& 1&   9  & 0&   -3.64&   -7.247&  2&   5 &  0&   -7.512&  -14.438\\
0.5& 1.5& 0   &-24.272& -49.645& 1&   9.5& 0&   -3.355&  -6.75&   2&   5.5& 0&   -6.768&  -13.002\\
0.5& 2  & 0   &-18.568 &-37.28 & 1&   10&  0&   -3.093&  -6.299&  2&   6&   0&   -6.133&  -11.795\\
0.5& 2.5& 0  & -15.038& -29.776&1 &  15& 0 &  -1.249&  -3.296&  2&   6.5& 0&   -5.584&  -10.765\\
0.5& 3  & 0   &-12.619& -24.726& 1&   20&  0&   -0.113&  -1.621&  2&   7&   0&   -5.103&  -9.875\\
0.5& 3.5& 0   &-10.848& -21.088& 1&   25&  0&   0.707&   -0.5 &   2&   7.5& 0&  -4.677&  -9.096\\
0.5& 4  & 0   &-9.486 & -18.337& 1&   30&  0&   1.343&   0.327&   2&   8&   0&   -4.296&  -8.408\\
0.5& 4.5& 0   &-8.401&  -16.179& 1.5& 0.5& 0.162&   -65.065& -100&    2&   8.5& 0&   -3.952&  -7.796\\
0.5& 5  & 0  & -7.512 & -14.439& 1.5& 1 &  0.002&   -35.218& -73.885& 2&   9&   0&   -3.64&  -7.247\\
0.5& 5.5& 0  & -6.768 & -13.003& 1.5& 1.5& 0  & -24.272& -49.645& 2&   9.5& 0&   -3.355&  -6.75\\
0.5& 6  & 0  & -6.134&  -11.796& 1.5& 2 &  0  & -18.568& -37.279& 2&   10&  0&   -3.092&  -6.299\\
0.5& 6.5& 0  & -5.585&  -10.766& 1.5& 2.5& 0  & -15.037& -29.776& 2&   15&  0&   -1.249&  -3.296\\
0.5& 7  & 0  & -5.104&  -9.876&  1.5& 3 &  0  & -12.619& -24.726& 2&   20&  0&   -0.112&  -1.62\\
0.5& 7.5& 0  & -4.678 & -9.097&  1.5& 3.5& 0  & -10.847& -21.088& 2&   25&  0&   0.707&   -0.499\\
0.5& 8  & 0  & -4.297&  -8.409&  1.5& 4 &  0  & -9.485&  -18.336& 2&   30&  0&   1.343&   0.328\\
0.5& 8.5& 0  & -3.953&  -7.797&  1.5& 4.5& 0  & -8.4 &   -16.178& 2.5& 0.5& 0.261&   -64.065& -100\\
0.5& 9  & 0  & -3.641&  -7.247&  1.5& 5 &  0  & -7.512&  -14.438& 2.5& 1&   0.015&   -35.149& -73.953\\
0.5& 9.5& 0  & -3.356&  -6.751&  1.5& 5.5& 0  & -6.768&  -13.002& 2.5& 1.5& 0.002&   -24.267& -49.649\\
0.5& 10 & 0  & -3.093&  -6.3&    1.5& 6 &  0  & -6.133&  -11.795& 2.5& 2 &  0 &  -18.567& -37.28\\
0.5& 15 & 0  & -1.25 &  -3.297&  1.5& 6.5& 0  & -5.584&  -10.766& 2.5& 2.5& 0&   -15.037& -29.776\\
0.5& 20 & 0 &  -0.114&  -1.622&  1.5& 7 &  0  & -5.103&  -9.875&  2.5& 3 &  0&   -12.619& -24.726\\
0.5& 25 & 0  & 0.705 &  -0.501&  1.5& 7.5& 0  & -4.677&  -9.096&  2.5& 3.5& 0&   -10.847& -21.088\\
0.5& 30 & 0  & 1.342 &  0.326&   1.5& 8 &  0  & -4.296&  -8.409&  2.5& 4 &  0&   -9.485&  -18.336\\
1 &  0.5& 0.113&   -65.563& -100&1.5& 8.5& 0  & -3.952&  -7.796&  2.5& 4.5& 0&   -8.4&    -16.178\\
1 &  1  & 0  & -35.223& -73.88&  1.5& 9 &  0  & -3.64&   -7.247&  2.5& 5 &  0&   -7.512&  -14.438\\
1 &  1.5& 0  & -24.272& -49.645& 1.5& 9.5& 0  & -3.355&  -6.75&   2.5& 5.5& 0&   -6.768&  -13.002\\
1 &  2  & 0  & -18.568& -37.279& 1.5& 10& 0  & -3.092&  -6.299&  2.5& 6  & 0&   -6.133&  -11.795\\
1 &  2.5& 0  & -15.037& -29.776& 1.5& 15&  0 &  -1.249&  -3.296&  2.5& 6.5& 0&   -5.584&  -10.765\\
1 &  3  & 0  & -12.619& -24.726& 1.5& 20&  0 &  -0.112&  -1.62&   2.5& 7 &  0&   -5.103&  -9.875\\
1 &  3.5& 0  & -10.847& -21.088& 1.5& 25&  0 &  0.707&   -0.499&  2.5& 7.5& 0&   -4.677&  -9.096\\
1 &  4  & 0  & -9.485&  -18.336& 1.5& 30&  0  & 1.343&   0.328&   2.5& 8  & 0 &  -4.296&  -8.408\\
1 &  4.5& 0  & -8.4  &  -16.179& 2  & 0.5& 0.212&   -64.565& -100&    2.5& 8.5& 0 &  -3.952&  -7.796\\
1 &  5  & 0  & -7.512&  -14.438& 2  & 1 &  0.005&   -35.201& -73.901& 2.5& 9&   0 &  -3.64&   -7.247\\
1 &  5.5& 0  & -6.768&  -13.002& 2  & 1.5& 0  & -24.27&  -49.646& 2.5& 9.5& 0 &  -3.355&  -6.75\\
1 &  6  & 0  & -6.133&  -11.796& 2  & 2 &  0  & -18.568& -37.28&  2.5& 10 & 0&   -3.092&  -6.299\\
1 &  6.5& 0  & -5.584&  -10.766& 2  & 2.5& 0  & -15.037& -29.776& 2.5& 15 & 0 & -1.249&  -3.296\\
1 &  7  & 0  & -5.103&  -9.875&  2  & 3 &  0  & -12.619& -24.726& 2.5& 20 & 0&   -0.112&  -1.62\\
1 &  7.5& 0  & -4.677&  -9.096&  2  & 3.5& 0  & -10.847& -21.088& 2.5& 25 & 0&   0.707&   -0.499\\
1 &  8  & 0 &  -4.296&  -8.409&  2  & 4&   0  & -9.485&  -18.336& 2.5& 30 & 0&   1.343&   0.328\\

\end{tabular}
\end{center}}
\newpage
\begin{center}
\begin{tabular}{ccccc|ccccc|ccccc} \\ \hline
$log\rho Y_{e}$& $T_{9}$ & $E_{f}$ & $\lambda^{ec}$ &
$\lambda^{pc}$ & $log\rho Y_{e}$& $T_{9}$ & $E_{f}$ &
$\lambda^{ec}$ & $\lambda^{pc}$ & $log\rho Y_{e}$& $T_{9}$ &
$E_{f}$ & $\lambda^{ec}$ & $\lambda^{pc}$   \\\hline
3&   0.5& 0.311&   -63.564& -100&    4&   0.5& 0.411&   -62.552& -100 &   5&   0.5& 0.522&   -61.433& -100\\
3&   1  & 0.046&   -34.995& -74.107& 4&   1 & 0.209&   -34.173& -74.929& 5&   1 &  0.413&   -33.142& -75.961\\
3&   1.5& 0.005&   -24.256& -49.661& 4&   1.5& 0.047&   -24.114& -49.802& 5&   1.5& 0.265&   -23.381& -50.536\\
3&   2  & 0.001 &   -18.565& -37.283& 4&   2 &  0.014&   -18.533& -37.314& 5&   2 &  0.128&   -18.244& -37.603\\
3&   2.5& 0.001 &   -15.036& -29.777& 4&   2.5& 0.006&   -15.025& -29.788& 5&   2.5& 0.062&   -14.912& -29.901\\
3&   3  & 0 &  -12.618& -24.726& 4&   3&   0.004&   -12.613& -24.732& 5&   3 &  0.035&   -12.559& -24.785\\
3&   3.5& 0 &  -10.847& -21.088& 4&   3.5& 0.002&   -10.844& -21.091& 5&   3.5& 0.023&   -10.814& -21.12\\
3&   4  & 0 &  -9.485&  -18.336& 4&   4 &  0.002&   -9.483&  -18.338& 5&   4 &  0.016&   -9.465&  -18.356\\
3&   4.5& 0 &  -8.4&    -16.179& 4&   4.5& 0.001&    -8.399&  -16.18&  5&   4.5& 0.012&   -8.387&  -16.192\\
3&   5 &  0 &  -7.512&  -14.438& 4&   5 &  0.001&    -7.511&  -14.439& 5&   5 &  0.009&   -7.502&  -14.447\\
3&   5.5& 0 &  -6.767&  -13.002& 4&   5.5& 0.001&    -6.767&  -13.003& 5&   5.5& 0.007&   -6.761&  -13.009\\
3&   6  & 0 &  -6.133&  -11.795& 4&   6 &  0.001&    -6.133&  -11.796& 5&   6&   0.006&   -6.128&  -11.8\\
3&   6.5& 0 &  -5.584&  -10.765& 4&   6.5& 0.001&    -5.584&  -10.766& 5&   6.5& 0.005&   -5.58&   -10.769\\
3&   7  & 0 &  -5.103&  -9.875&  4&   7 &  0&   -5.103&  -9.875&  5&   7&   0.004&   -5.1&    -9.878\\
3&   7.5& 0 &  -4.677&  -9.096&  4&   7.5& 0&   -4.677&  -9.096&  5&   7.5& 0.004&   -4.674&  -9.099\\
3&   8  & 0 &  -4.296&  -8.408&  4&   8  & 0&   -4.296&  -8.409&  5&   8&   0.003&   -4.294&  -8.41\\
3&   8.5& 0 &  -3.952&  -7.796&  4&   8.5& 0&   -3.952&  -7.796&  5&   8.5& 0.003&   -3.95&   -7.798\\
3&   9  & 0 &  -3.64&   -7.246&  4&   9 &  0&   -3.64&   -7.247&  5&  9&   0.002&   -3.639& -7.248\\
3&   9.5& 0 &  -3.355&  -6.75&   4&   9.5& 0&   -3.355&  -6.75&   5&   9.5& 0.002&   -3.353&  -6.751\\
3&   10 & 0 &  -3.092&  -6.299&  4&   10&  0&   -3.092&  -6.299&  5&   10&  0.002&   -3.091&  -6.3\\
3&   15 & 0  & -1.248&  -3.296&  4&   15&  0&   -1.248&  -3.296&  5&   15&  0.001&    -1.248&  -3.296\\
3&   20 & 0 &  -0.112&  -1.62 &  4&   20&  0&   -0.112&  -1.62 &  5&   20&  0&   -0.112&  -1.62\\
3&   25 & 0 &  0.707&   -0.499&  4&   25&  0&   0.707&   -0.499&  5&   25&  0&   0.707&   -0.499\\
3&   30 & 0 &  1.344&   0.328&   4&   30&  0&   1.344&   0.328&   5&   30&  0&   1.344&   0.328\\
3.5& 0.5& 0.361&   -63.061& -100&    4.5& 0.5& 0.464&   -62.023& -100&    5.5& 0.5& 0.598&   -60.675& -100\\
3.5& 1 &  0.115&   -34.647& -74.455& 4.5& 1  & 0.309&   -33.667& -75.435& 5.5& 1&   0.528&   -32.563& -76.54\\
3.5& 1.5& 0.015&   -24.221& -49.695& 4.5& 1.5& 0.13&    -23.836& -50.08&  5.5& 1.5& 0.423&   -22.852& -51.066\\
3.5& 2 &  0.004&   -18.557& -37.29&  4.5& 2 &  0.044&   -18.458& -37.389& 5.5& 2&   0.295&   -17.824& -38.024\\
3.5& 2.5& 0.002&   -15.033&-29.78&  4.5& 2.5& 0.02&    -14.997& -29.816& 5.5& 2.5& 0.18&    -14.674& -30.139\\
3.5& 3 &  0.001&    -12.617& -24.728& 4.5& 3&   0.011&   -12.6&   -24.745& 5.5& 3&   0.11&    -12.435& -24.91\\
3.5& 3.5& 0.001&    -10.846& -21.089& 4.5& 3.5& 0.007&   -10.837& -21.098& 5.5& 3.5& 0.072&   -10.744& -21.191\\
3.5& 4 &  0.001&    -9.485&  -18.337& 4.5& 4  & 0.005&   -9.479&  -18.342& 5.5& 4&   0.05&    -9.422&  -18.4\\
3.5& 4.5& 0&   -8.4&    -16.179& 4.5& 4.5& 0.004&  -8.396&  -16.183& 5.5& 4.5& 0.038&   -8.358&  -16.22\\
3.5& 5  & 0&   -7.511&  -14.438& 4.5& 5 &  0.003&   -7.509&  -14.441& 5.5& 5&   0.029&   -7.482&  -14.467\\
3.5& 5.5& 0&   -6.767&  -13.002& 4.5& 5.5& 0.002&   -6.765&  -13.004& 5.5& 5.5& 0.023&  -6.746&  -13.023\\
3.5& 6  & 0&   -6.133&  -11.796& 4.5& 6 &  0.002&   -6.132&  -11.797& 5.5& 6&   0.019&   -6.117&  -11.811\\
3.5& 6.5& 0&   -5.584&  -10.766& 4.5& 6.5& 0.002&   -5.583&  -10.767& 5.5& 6.5& 0.016&   -5.572&  -10.778\\
3.5& 7 &  0&   -5.103&  -9.875&  4.5& 7 &  0.001&   -5.102&  -9.876&  5.5& 7&   0.013&   -5.093&  -9.885\\
3.5& 7.5& 0&   -4.677&  -9.096&  4.5& 7.5& 0.001&    -4.676&  -9.097&  5.5& 7.5& 0.012&   -4.669& -9.104\\
3.5& 8 &  0&   -4.296&  -8.408&  4.5& 8  & 0.001&    -4.295&  -8.409&  5.5& 8&   0.01&    -4.289&  -8.415\\
3.5& 8.5& 0&   -3.952&  -7.796&  4.5& 8.5& 0.001&    -3.952&  -7.797&  5.5& 8.5& 0.009&   -3.947& -7.801\\
3.5& 9 &  0&   -3.64&   -7.247&  4.5& 9  & 0.001&    -3.64&   -7.247&  5.5& 9&   0.008&   -3.636&  -7.251\\
3.5& 9.5& 0&   -3.355&  -6.75&   4.5& 9.5& 0.001&    -3.354&  -6.75&   5.5& 9.5& 0.007&   -3.351&  -6.754\\
3.5& 10&  0&   -3.092&  -6.299&  4.5& 10&  0.001&    -3.092&  -6.299&  5.5& 10&  0.006&   -3.089&  -6.302\\
3.5& 15&  0&   -1.248&  -3.296&  4.5& 15&  0&   -1.248&  -3.296&  5.5& 15&  0.003&   -1.248&  -3.296\\
3.5& 20&  0&   -0.112&  -1.62 &  4.5& 20&  0&   -0.112&  -1.62&   5.5& 20&  0.001&    -0.112&  -1.621\\
3.5& 25&  0&   0.707&   -0.499&  4.5& 25&  0&   0.707&   -0.499&  5.5& 25&  0.001&    0.707&   -0.499\\
3.5& 30&  0&   1.344&   0.328&   4.5& 30&  0&   1.344&   0.328&   5.5& 30&  0.001&    1.344&   0.328\\
\end{tabular}
\end{center}
\newpage
\begin{center}
\begin{tabular}{ccccc|ccccc|ccccc} \\ \hline
$log\rho Y_{e}$& $T_{9}$ & $E_{f}$ & $\lambda^{ec}$ &
$\lambda^{pc}$ & $log\rho Y_{e}$& $T_{9}$ & $E_{f}$ &
$\lambda^{ec}$ & $\lambda^{pc}$ & $log\rho Y_{e}$& $T_{9}$ &
$E_{f}$ & $\lambda^{ec}$ & $\lambda^{pc}$   \\\hline
6&   0.5& 0.713&   -59.518& -100&    7&   0.5& 1.217&   -56.24&  -100&    8&   0.5& 2.444&   -45.14&  -100\\
6&   1  & 0.672&   -31.846& -77.264& 7&   1 &  1.2& -29.578& -79.927& 8&   1&   2.437&   -23.824& -86.158\\
6&   1.5& 0.604&   -22.248& -51.673& 7&   1.5& 1.173&   -20.447& -53.584& 8&   1.5& 2.424&   -16.495& -57.788\\
6&   2 &  0.512&   -17.279& -38.571& 7&   2 &  1.133&   -15.756& -40.136& 8&   2&   2.406&   -12.716& -43.342\\
6&   2.5& 0.405&   -14.223& -30.592& 7&  2.5& 1.083&   -12.877& -31.959& 8&   2.5& 2.383&   -10.375& -34.58\\
6&   3 &  0.299&   -12.117& -25.228& 7&   3 &  1.021&   -10.916& -26.442& 8&   3&   2.355&   -8.766& -28.682\\
6&   3.5& 0.214&   -10.539& -21.396& 7&   3.5& 0.95&    -9.488&  -22.456& 8&   3.5& 2.322&   -7.582&  -24.431\\
6&   4 &  0.156&   -9.289&  -18.532& 7&   4 &  0.871&   -8.395&  -19.433& 8&   4&   2.283&   -6.669&  -21.213\\
6&   4.5& 0.117&   -8.269&  -16.31&  7&   4.5& 0.785&   -7.525&  -17.058& 8&   4.5& 2.24&    -5.94&   -18.687\\
6&   5  & 0.091&   -7.42&   -14.53&  7 &  5 & 0.698&   -6.811&  -15.142& 8&   5&   2.192&   -5.342&  -16.648\\
6&   5.5& 0.073&   -6.701&  -13.069& 7&   5.5& 0.613&   -6.209&  -13.564& 8&   5.5& 2.139&   -4.841&  -14.962\\
6&   6 &  0.06&    -6.083&  -11.846& 7&   6  & 0.534&   -5.686&  -12.244& 8&   6 &  2.081&   -4.414&  -13.543\\
6&   6.5& 0.05&    -5.546&  -10.804& 7&   6.5& 0.465&   -5.226&  -11.126& 8&   6.5& 2.019&   -4.043&  -12.331\\
6&   7 &  0.042&   -5.073&  -9.905&  7&   7  & 0.404&   -4.814&  -10.166& 8&   7&   1.952&   -3.719&  -11.28\\
6&   7.5& 0.036&   -4.652&  -9.121&  7&   7.5& 0.353&   -4.441&  -9.333&  8&   7.5& 1.882&   -3.431&  -10.361\\
6&   8 &  0.032&   -4.276&  -8.428&  7&   8  & 0.31&    -4.101&  -8.604&  8&   8 &  1.808&   -3.173&  -9.548\\
6&   8.5& 0.028&   -3.936&  -7.812&  7&   8.5& 0.274&   -3.79&   -7.959&  8&   8.5& 1.732&   -2.94&   -8.823\\
6&   9 &  0.025&  -3.626&  -7.26&   7&  9&   0.244&   -3.504&  -7.383&  8 &  9 &  1.653&   -2.727&  -8.172\\
6&   9.5& 0.022&   -3.343&  -6.762&  7&   9.5& 0.218&   -3.24&   -6.866&  8&   9.5& 1.574&   -2.531&  -7.585\\
6&   10&  0.02 &   -3.082&  -6.309&  7&   10&  0.196&   -2.994&  -6.398&  8&   10&  1.493&   -2.35&   -7.051\\
6&   15&  0.009&   -1.246&  -3.298&  7&   15&  0.085&   -1.22&   -3.324&  8&   15&  0.817&   -0.978&  -3.57\\
6&   20&  0.005&   -0.111&  -1.621&  7&   20&  0.047&   -0.1&    -1.632&  8&   20&  0.47&    0.004&   -1.738\\
6&   25&  0.003&   0.708&   -0.5&    7&   25&  0.03&    0.713&   -0.505&  8&   25&  0.301&   0.767&   -0.56\\
6&  30 & 0.002 &  1.344&   0.328&  7&   30&  0.021&   1.347&   0.324&   8&   30&  0.209&   1.378&   0.293\\
6.5& 0.5& 0.905&   -57.811& -100&    7.5& 0.5& 1.705&   -52.593& -100&    8.5& 0.5& 3.547&   -34.019& -100\\
6.5& 1 &  0.88&    -30.857& -78.31&  7.5& 1 &  1.693&   -27.533& -82.412& 8.5& 1 &  3.542&   -18.252& -91.73\\
6.5& 1.5& 0.837&   -21.483& -52.456& 7.5& 1.5& 1.675&   -18.944& -55.271& 8.5& 1.5& 3.534&   -12.769& -61.517\\
6.5& 2 &  0.777&   -16.621& -39.237& 7.5& 2 &  1.648&   -14.55&  -41.433& 8.5& 2&   3.521&   -9.916&  -46.153\\
6.5& 2.5& 0.701&  -13.63&  -31.189& 7.5& 2.5& 1.614&   -11.857& -33.03&  8.5& 2.5& 3.506&   -8.134&  -36.843\\
6.5& 3 &  0.612&   -11.594& -25.754& 7.5& 3&   1.573&   -10.022& -27.368& 8.5& 3&   3.487&   -6.898&  -30.583\\
6.5& 3.5& 0.517&   -10.105& -21.832& 7.5& 3.5& 1.524&   -8.683&  -23.282& 8.5& 3.5& 3.464&  -5.978&  -26.076\\
6.5& 4  & 0.424&   -8.953&  -18.87&  7.5& 4 &  1.468&   -7.658&  -20.186& 8.5& 4 &  3.438&   -5.259&  -22.668\\
6.5& 4.5& 0.343&   -8.018&  -16.562& 7.5& 4.5& 1.405&   -6.843&  -17.752& 8.5& 4.5& 3.408&   -4.677&  -19.996\\
6.5& 5 &  0.277&   -7.233&  -14.717& 7.5& 5 &  1.336&   -6.178&  -15.785& 8.5& 5 &  3.375&   -4.194&  -17.84\\
6.5& 5.5& 0.226&  -6.561&  -13.209& 7.5& 5.5& 1.262&   -5.622&  -14.158& 8.5& 5.5& 3.339&   -3.784&  -16.062\\
6.5& 6 &  0.187&   -5.977&  -11.952& 7.5& 6 & 1.183&   -5.148&  -12.789& 8.5& 6&   3.299&   -3.431&  -14.566\\
6.5& 6.5& 0.157&   -5.463&  -10.887& 7.5& 6.5& 1.101&   -4.738&  -11.619& 8.5& 6.5& 3.256&   -3.122&  -13.29\\
6.5& 7 &  0.134&   -5.007&  -9.971&  7.5& 7 &  1.018&   -4.376&  -10.608& 8.5& 7&   3.209&  -2.849&  -12.185\\
6.5& 7.5& 0.115&   -4.6&    -9.173&  7.5& 7.5& 0.937&   -4.053&  -9.725&  8.5& 7.5& 3.159&   -2.606&  -11.219\\
6.5&8&   0.1& -4.233&  -8.471&  7.5& 8&   0.858&   -3.76&   -8.949&  8.5& 8&   3.106&   -2.386&  -10.365\\
6.5& 8.5& 0.088&   -3.9&    -7.848&  7.5& 8.5& 0.783&   -3.492&  -8.26&   8.&5 8.5& 3.05&    -2.187&  -9.604\\
6.5& 9 &  0.078&   -3.597&  -7.29&   7.5& 9 &  0.714&   -3.244&  -7.646&  8.5& 9&   2.99&    -2.005&  -8.921\\
6.5& 9.5& 0.069&   -3.318&  -6.787&  7.5& 9.5& 0.651&   -3.012&  -7.095&  8.5& 9.5& 2.928&   -1.838&  -8.303\\
6.5& 10&  0.062&   -3.061&  -6.33&   7.5& 10&  0.593&   -2.796&  -6.598&  8.5& 10&  2.863&   -1.683&  -7.741\\
6.5&15&  0.027 &  -1.239&  -3.305&  7.5& 15 & 0.268&   -1.159&  -3.386&  8.5& 15&  2.109&   -0.554&  -4.004\\
6.5& 20&  0.015&   -0.109&  -1.624&  7.5& 20&  0.15&    -0.075&  -1.658&  8.5& 20&  1.402&   0.235&   -1.973\\
6.5& 25&  0.01 &   0.709 &  -0.501&  7.5& 25&  0.095&   0.726&   -0.518&  8.5& 25&  0.936&   0.892&   -0.687\\
6.5& 30&  0.007 &  1.345&   0.327&   7.5& 30&  0.066&   1.355&   0.317&   8.5& 30&  0.656&   1.452&   0.218\\
\end{tabular}
\end{center}
\newpage
\begin{center}
\begin{tabular}{ccccc|ccccc|ccccc} \\ \hline
$log\rho Y_{e}$& $T_{9}$ & $E_{f}$ & $\lambda^{ec}$ &
$\lambda^{pc}$ & $log\rho Y_{e}$& $T_{9}$ & $E_{f}$ &
$\lambda^{ec}$ & $\lambda^{pc}$ & $log\rho Y_{e}$& $T_{9}$ &
$E_{f}$ & $\lambda^{ec}$ & $\lambda^{pc}$   \\\hline
9&   0.5& 5.179&   -17.57&  -100&    9.5& 8.5& 7.351&   0.199&   -12.155& 10.5&    4.5& 16.28&   2.986&  -34.412\\
9&   1&   5.176&   -10.019& -99.963& 9.5& 9&   7.323&   0.258&   -11.347& 10.5&    5&   16.273&  2.983&   -30.841\\
9&   1.5& 5.17&    -7.271&  -67.015& 9.5&9.5& 7.293&   0.315&   -10.619& 10.5&    5.5& 16.265&  2.98 &   -27.907\\
9&   2&   5.162&   -5.783&  -50.286& 9.5& 10&  7.261&   0.373&   -9.958&  10.5&    6&   16.256&  2.977&   -25.451\\
9&   2.5& 5.151&   -4.819&  -40.16&  9.5& 15&  6.86&    0.917&   -5.6&    10.5&    6.5& 16.247&  2.974&   -23.363\\
9&   3&   5.138&   -4.128& -33.357& 9.5&20&  6.307&   1.392&   -3.209&  10.5&    7&   16.237&  2.971 &  -21.565\\
9&   3.5& 5.122&   -3.599&  -28.464& 9.5& 25&  5.624&   1.792&   -1.632&  10.5&    7.5& 16.226&  2.968&   -20\\
9&   4&   5.105&   -3.175&  -24.768& 9.5& 30&  4.859&   2.131&   -0.487&  10.5&    8&   16.214&  2.966&   -18.623\\
9&   4.5& 5.085&   -2.824&  -21.873& 10&  0.5& 11.118&  1.613&   -100&    10.5&    8.5& 16.202&  2.964&   -17.403\\
9&   5&   5.062&   -2.525&  -19.54&  10&  1&   11.116&  1.615&   -100&    10.5&    9&   16.189&  2.962&   -16.312\\
9&   5.5& 5.037&   -2.266&  -17.618& 10&  1.5& 11.113&  1.618&   -86.985& 10.5&    9.5& 16.175&  2.961&   -15.331\\
9&   6&  5.01&    -2.037&  -16.004& 10&  2&   11.11&   1.622&   -65.275& 10.5&    10&  16.16&   2.96&    -14.444\\
9&   6.5& 4.98&    -1.833&  -14.627& 10& 2.5& 11.105&  1.626&   -52.162& 10.5&    15&  15.973&  3.005&   -8.662\\
9&   7&   4.948&   -1.648&  -13.438& 10&  3&   11.099&  1.631&   -43.371& 10.5&    20&  15.711&  3.152&   -5.579\\
9&   7.5& 4.914&   -1.479&  -12.398& 10&  3.5& 11.091&  1.635&   -37.059& 10.5&    25&  15.375&  3.353&   -3.598\\
9&   8&   4.878&   -1.324&  -11.481& 10&  4 &  11.083&  1.64&    -32.301& 10.5&    30&  14.965&  3.555&   -2.185\\
9&   8.5& 4.839&   -1.181& -10.665& 10&  4.5& 11.074&  1.645&   -28.581& 11&  0.5& 23.934&  4.131&   -100\\
9&   9&   4.797&   -1.048&  -9.933&  10&  5&   11.063&  1.649&   -25.59&  11&  1&   23.933&  4.131&   -100\\
9&   9.5& 4.754&   -0.923&  -9.272&  10&  5.5& 11.052&  1.654&   -23.13&  11&  1.5& 23.932&  4.131&   -100\\
9&   10&  4.708&   -0.806&  -8.671&  10 & 6&   11.039&  1.66&    -21.068& 11&  2 &  23.93&   4.131&   -97.582\\
9&   15&  4.131&   0.091&  -4.683&  10&  6.5& 11.025&  1.666&   -19.314& 11&  2.5& 23.928&  4.131&   -78.013\\
9&   20&  3.39&    0.717&   -2.474&  10& 7&   11.011&  1.672&   -17.802& 11&  3 &  23.925&  4.13 &   -64.919\\
9&   25&  2.621&   1.222&   -1.027&  10&  7.5& 10.995&  1.68&    -16.484& 11&  3.5& 23.922&  4.128&   -55.534\\
9&   30&  1.973&   1.668&   -0.003&  10&  8&   10.978&  1.688&   -15.324& 11&  4&   23.918&  4.126&   -48.472\\
9.5& 0.5& 7.583&   -0.75&   -100&    10& 8.5& 10.959&  1.697&   -14.294& 11&  4.5& 23.913&  4.122&   -42.961\\
9.5& 1&   7.581&   -0.726&  -100&    10&  9&   10.94&   1.707&   -13.373& 11&  5 &  23.908&  4.118&   -38.538\\
9.5& 1.5& 7.577&   -0.688&  -75.102& 10& 9.5& 10.92&   1.718&   -12.543& 11&  5.5& 23.903& 4.113&   -34.906\\
9.5& 2&   7.571&   -0.638&  -56.359& 10&  10&  10.898&  1.731&   -11.792& 11&  6 &  23.897&  4.107&   -31.869\\
9.5& 2.5& 7.564&   -0.579&  -45.024& 10&  15&  10.624&  1.927&   -6.865&  11&  6.5& 23.891&  4.102&   -29.29\\
9.5& 3&   7.555&   -0.514&  -37.418& 10&  20&  10.241&  2.216&   -4.2&    11&  7 &  23.884&  4.096&   -27.071\\
9.5& 3.5& 7.545&   -0.447&  -31.952& 10&  25&  9.751&   2.514&   -2.464&  11&  7.5& 23.877&  4.091&   -25.141\\
9.5& 4&   7.532&   -0.378&  -27.827& 10&  30&  9.163&   2.783&   -1.21&   11&  8&   23.869&  4.085&   -23.445\\
9.5& 4.5& 7.519&   -0.308&  -24.599& 10.5&    0.5& 16.31&   2.987&   -100&    11&  8.5& 23.86&   4.08&    -21.944\\
9.5& 5&   7.503&   -0.24&   -22.001& 10.5&    1&   16.309&  2.987&   -100&    11&  9&   23.851&  4.075&   -20.603\\
9.5& 5.5& 7.486&   -0.173&  -19.862& 10.5&    1.5& 16.307&  2.988&   -100&    11&  9.5& 23.842&  4.07&    -19.399\\
9.5& 6&   7.468&   -0.108&  -18.068&10.5&   2&   16.304&  2.989&   -78.366& 11&  10&  23.832&  4.066&   -18.31\\
9.5& 6.5& 7.448&   -0.044&  -16.54&  10.5&    2.5& 16.301&  2.989&   -62.638& 11&  15&  23.704&  4.055&   -11.26\\
9.5& 7 &  7.426&   0.018&   -15.221& 10.5&    3&   16.297&  2.99&    -52.104& 11&  20&  23.526&  4.127&   -7.548\\
9.5& 7.5& 7.403&   0.079&   -14.07&  10.5&   3.5& 16.292&  2.989&   -44.548& 11&  25& 23.296&  4.26 &   -5.194\\
9.5& 8&   7.378&   0.14&    -13.056& 10.5&    4&   16.286&  2.988&
-38.857& 11&  30&  23.016&  4.408&   -3.537\\\hline

\end{tabular}
\end{center}

\end{document}